\documentclass[reprint,aps,pra,superscriptaddress]{revtex4-1}
\usepackage{amssymb,amsmath}
\usepackage{graphicx}
\usepackage{dcolumn}
\usepackage{multirow}
\usepackage[usenames,dvipsnames]{color}
\usepackage[english]{babel}
\usepackage[caption=false]{subfig} 
\usepackage{gensymb}

\newcommand{\ket}[1]{\ensuremath{|{#1}\rangle}}

\definecolor{darkgreen}{rgb}{0,0.5,0}

\begin{document}

\def\simlt{\mathrel{\lower .3ex \rlap{$\sim$}\raise .5ex \hbox{$<$}}}

\title{\textbf{\fontfamily{phv}\selectfont 
Microwave-driven coherent operation of a semiconductor quantum dot charge qubit }}
\author{Dohun Kim}
\author{D. R. Ward}
\author{C. B. Simmons}
\affiliation{Department of Physics, University of Wisconsin-Madison, Madison, WI 53706}
\author{John King Gamble}
\author{Robin Blume-Kohout}
\author{Erik Nielsen}
\affiliation{Sandia National Laboratories, Albuquerque, NM 87185, USA}
\author{D. E. Savage}
\author{M. G. Lagally}
\affiliation{Department of Materials Science and Engineering, University of Wisconsin-Madison, Madison, WI 53706, USA}
\author{Mark Friesen}
\author{S. N. Coppersmith}
\author{M. A. Eriksson}
\affiliation{Department of Physics, University of Wisconsin-Madison, Madison, WI 53706}

\maketitle

\textbf{A most intuitive realization of a qubit is a single electron charge sitting at two well-defined positions, such as the left and right sides of a double quantum dot.  This qubit is not just simple but also has the potential for high-speed operation, because of the strong coupling of electric fields to the electron.  However, charge noise also couples strongly to this qubit, resulting in rapid dephasing at nearly all operating points, with the exception of one special ``sweet spot.''  Fast dc voltage pulses have been used to manipulate semiconductor charge qubits~\cite{Cao:2013p1401,Shinkai:2009p056802,Hayashi:2003p226804,Nakamura:1999p786,Petersson:2010p246804,Dovzhenko:2011p161802,Shi:2013p075416},
but these previous experiments did not achieve high-fidelity control, because dc gating requires excursions away from the sweet spot.
Here, by using resonant ac microwave driving, we achieve coherent manipulation of a semiconductor charge qubit, demonstrating an Rabi frequency of up to 2~GHz, a value approaching the intrinsic qubit frequency of 4.5~GHz. $Z$-axis rotations of the qubit are well-protected at the sweet spot, and by using ac gating, we demonstrate the same protection for rotations about arbitrary axes in the $X$-$Y$ plane of the qubit Bloch sphere. We characterize operations on the qubit using two independent tomographic approaches: standard process tomography~\cite{NielsenBook, Chow:2009p090502} and a newly developed method known as gate set tomography~\cite{BlumeKohout:2013p1310.4492}.  Both approaches show that this qubit can be operated with process fidelities greater than 86\% with respect to a universal set of unitary single-qubit operations.}

\begin{figure*}
\includegraphics[width=0.95\textwidth]{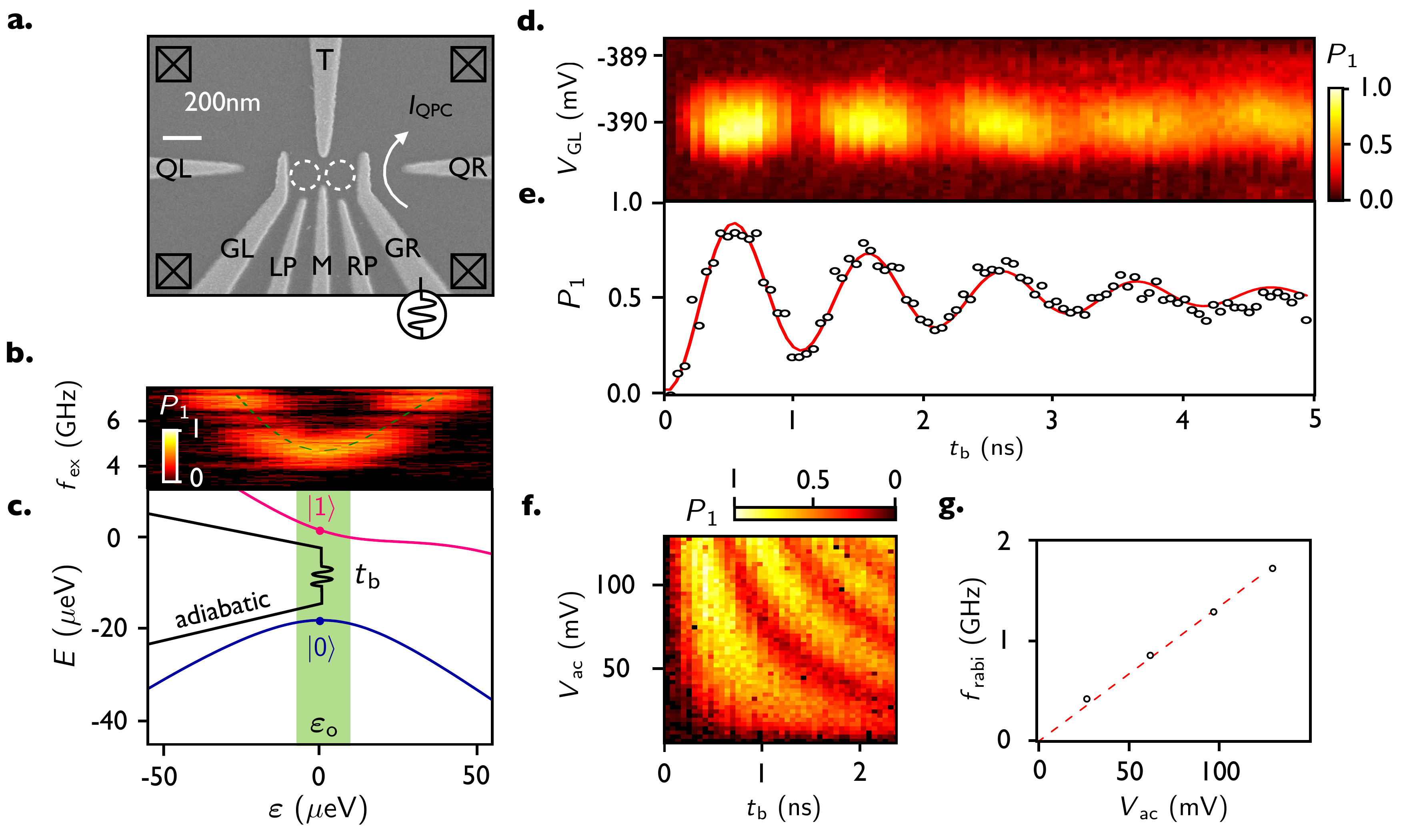}
\caption{
\textbf{Si/SiGe quantum dot device, qubit spectroscopy, and coherent Rabi oscillation measurements. a}, 
SEM image of a device lithographically identical to the one used in the experiment, with the locations of the double dot indicated by white dashed circles~\cite{Shi:2014p3020}. The current through the quantum point contact (QPC) $I_\text{QPC}$ is used for charge sensing via a measurement of its change when microwave component is added to a voltage sequence applied to gate GR.
\textbf{b-c}, Qubit energy levels and microwave spectroscopy. \textbf{b}, Probability $P_1$ of the state to be $\ket{1}$ at the end of the driving sequence as a function of detuning $\varepsilon$ and excitation frequency $f_\text{ex}$ of the microwave applied to gate GR. For this experiment, the repetition rate of the driving sequence is $\approx$ 15~MHz and the microwave pulses are of duration 10\ ns. The base value of the detuning $\varepsilon_{r}$ used for both readout and initialization is $\varepsilon=-160~\mu eV$. $P_1$ is large when the microwave pulse is resonant, so that it excites the system from the ground state to the excited state. Dashed green curve shows fit to calculated energy difference between ground state and lowest-energy excited state of the three-level model of Ref.~\cite{Kim:2014nature} (see also Supplementary Information S2). The third level of higher energy affects the dispersion of energy levels shown, but is otherwise unimportant because its occupation is negligible for all experiments shown.
\textbf{c}, Diagram of the calculated energy levels $E$ versus detuning $\varepsilon$, including the ground states of the $(2,1)$ and $(1,2)$ charge configuration, $|L\rangle$ and $|R\rangle$ respectively, and logical states $|0\rangle=(|L\rangle+|R\rangle)/\sqrt{2}$ and $|1\rangle=(|L\rangle-|R\rangle)/\sqrt{2}$. Black solid line inset: pulse sequence used for Rabi oscillation and spectroscopy measurements. 
\textbf{d-e}, Rabi oscillation. \textbf{d}, $P_1$ as a function of the voltage $V_\text{GL}$ and microwave pulse duration $t_\text{b}$ with $f_\text{ex}$ = 4.54 GHz and excitation amplitude $V_\text{ac}$ = 70 mV. 
\textbf{e}, Line-cut of $P_1$ near $V_\text{GL} = $-390 mV, showing $\approx$ 1 GHz coherent Rabi oscillations. Red solid curve shows a fit to an exponentially damped sine wave with best fit parameter $T_{2}^*$ = 1.5 ns.
\textbf{f-g}, Dependence of the Rabi oscillation frequency on the microwave amplitude. \textbf{f}, $P_1$ as a function of $V_\text{ac}$ and $t_\text{b}$ with $f_\text{ex}$ = 4.54 GHz, which demonstrates that the Rabi oscillation frequency $f_\text{Rabi}$ increases as the amplitude of the microwave driving is increased.
\textbf{g}, Rabi oscillation frequency $f_\text{Rabi}$ as a function of $V_\text{ac}$ with fixed $f_\text{ex}$ = 4.54 GHz. The good agreement of a linear fit (red dashed line) to the data is strong evidence that the measured oscillations are indeed Rabi oscillations, with the Rabi frequency proportional to the driving amplitude.}

\label{fig1} 
\end{figure*}

Coherent control of qubits with resonant microwaves plays an essential role in realizing precise single~\cite{Vion:2002p886,Chow:2009p090502, Chow:2010p040305} and two~\cite{ Fedorov:2012p170} qubit gates in solid state quantum computing architectures~\cite{ DiVincenzo:2000p1536}. In electrically-controlled quantum dots, driven coherent oscillations have been demonstrated in spin-based qubits using electron spin resonance \cite{Koppens:2006p766, Koppens:2008p236802, Pla:2012p489}, electric-dipole spin resonance \cite{Nowack:2007p1430, Berg:2013p066806, Petersson:2012p380, Kawakami2014unpublished}, or resonant exchange gates  \cite{Medford:2013p050501}, with typical rotation rates (Rabi frequencies) on the order of 1 to 100~MHz. Here we demonstrate fast coherent operation and full control on the Bloch sphere of a charge qubit in a double quantum dot in a silicon/silicon-germanium (Si/SiGe) heterostructure.
 
The charge qubit we study is formed by three electrons in a Si/SiGe double quantum dot (Fig.~1a)~\cite{ Shi:2013p075416, Shi:2014p3020}.  Fig.~1c shows the qubit energy level diagram as a function of the detuning $\varepsilon$, which is controlled by gate GL or GR (Fig.~1a).  The states $\ket{2,1}=\ket{L}$ and $\ket{1,2}=\ket{R}$ are the ground states of the system at negative and positive $\varepsilon$, respectively, and these states anticross at $\varepsilon=0$.  A state $\ket{R}_\text{e}$, of higher energy ($>$14 GHz with respect to $\ket{1}$ near $\varepsilon=0$ \cite{Kim:2014nature}) and not visible in the energy range shown in the figure, affects the dispersion of energy levels shown, but is otherwise unimportant because its occupation is negligible for the resonant driving demonstrated here. 

We focus on detunings near $\varepsilon\approx 0$, where an avoided crossing is formed between states $|L\rangle$ and $|R\rangle$ with tunnel coupling strength $\Delta_{1}$. The logical qubit states are the energy eigenstates at $\varepsilon=0$: $|0\rangle=(|L\rangle+|R\rangle)/\sqrt{2}$ and $|1\rangle=(|L\rangle-|R\rangle)/\sqrt{2}$. These states can be adiabatically evolved to states $|L\rangle$ and $|R\rangle$ at moderate negative $\varepsilon$, enabling robust readout of $P_1$, the probability of occupation of $\ket{1}$ at the end of qubit manipulation, performed by monitoring the current $I_\text{QPC}$ through the charge-sensing quantum point contact (Fig.~1a). 
 
We first characterize the system spectroscopically. 
As shown schematically by the black solid line in Fig.\ 1c, state $\ket{L}$ is first prepared by waiting longer than the charge relaxation time ($T_{1} = 23.5$~ns) at a detuning appropriate for initialization and readout, $\varepsilon_{r}\approx -160~\mu eV$ \cite{Kim:2014nature}.  We then ramp the gate voltage over a time of 4~ns to change the detuning to a value near $\varepsilon=0$, adiabatically evolving state $\ket{L}$ to state $\ket{0}$, completing the qubit initialization. At this detuning, a 10~ns microwave burst is then applied to gate GR.
When the microwave frequency is resonant with the splitting between the qubit energy levels, excitation occurs from $\ket{0}$ to $\ket{1}$. 
The resulting probability of state $\ket{1}$ is measured by ramping the detuning adiabatically over approximately 2~ns back to $\varepsilon_\text{r}$, transforming
$\ket{0}$ to $\ket{L}$ and $\ket{1}$ to $\ket{R}$,
and measuring the change in $I_\text{QPC}$~\cite{Kim:2014nature}.
The details of the measurement procedure and conversion to probability are presented in the Supplementary Information S1. Fig.\ 1b shows 
the resulting spectroscopy of the qubit energy levels. We find a good agreement between the spectroscopic measurement with the calculated lowest-energy excitation spectrum (green dashed curve) with Hamiltonian parameters measured in Ref.~\cite{Kim:2014nature}. Because the low-lying excited state on the right dot $\ket{R}_\text{e}$ affects energy dispersion, we measure a minimum resonant frequency of approximately 4.5~GHz near $\varepsilon=0$, typically lower than the tunnel coupling $2\Delta_\text{1}/h$ = 5.2 GHz used for calculation. At this sweet spot $\partial E / \partial \varepsilon=0$, and the energy levels are first order insensitive to detuning noise~\cite{Petersson:2010p246804, Shi:2013p075416, Medford:2013p050501}.    
 
\begin{figure*}
\includegraphics[width=0.95\textwidth]{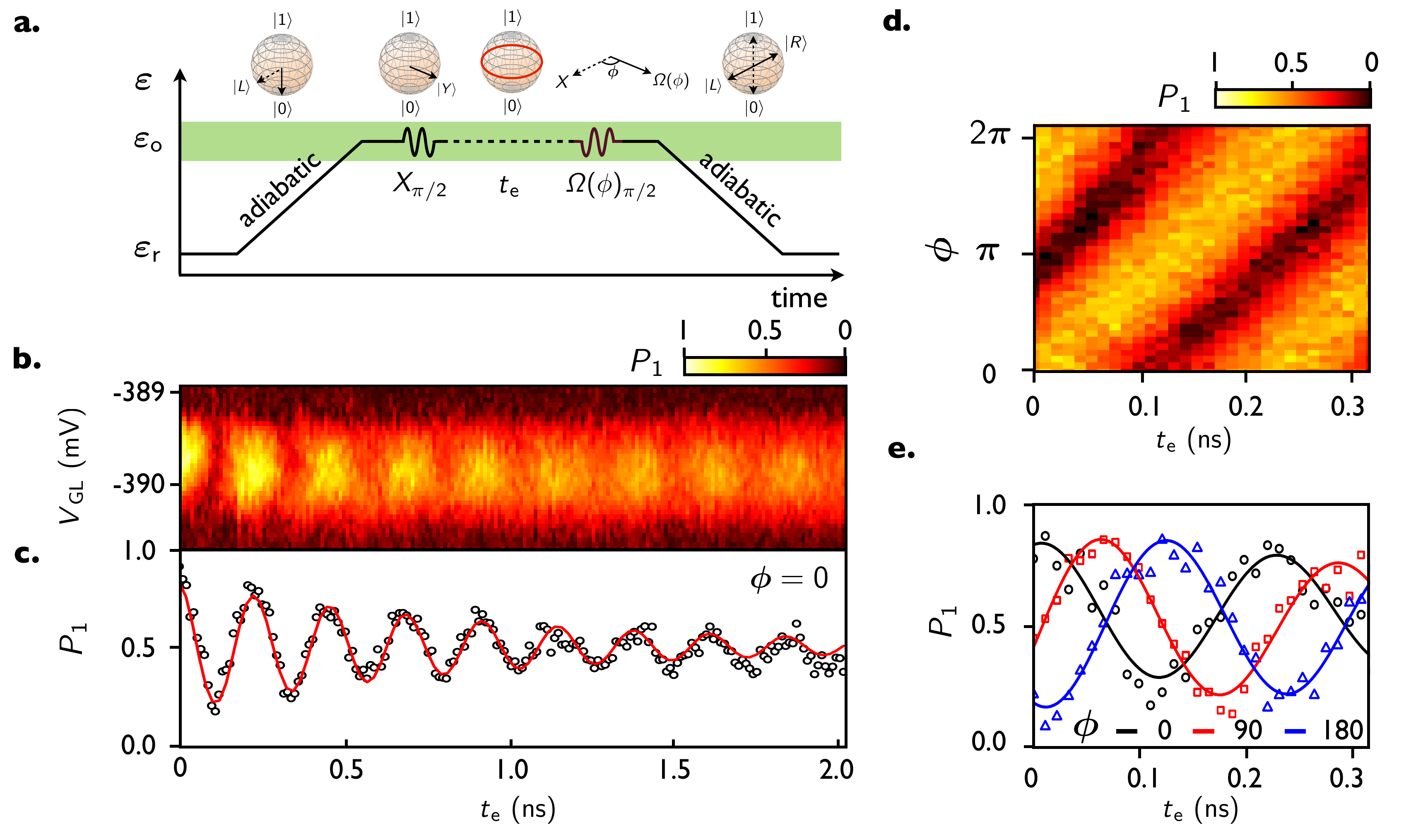}
\caption{
\textbf{Ramsey fringes and demonstration of three axes control of ac-gated charge qubit. a}, Schematic of the pulse sequences used to perform universal control of the qubit. Both the delay $t_\text{e}$ and the phase $\phi$ of the second microwave pulse are varied in the experiment.
\textbf{b-c}, Experimental measurement of Z-axis rotation. In a Ramsey fringe (Z-axis rotation) measurement, the first $X_\text{$\pi$/2}$ gate rotates the Bloch vector onto the $X$-$Y$ plane, and the second $X_\text{$\pi$/2}$ gate ($\phi$ = 0) is delayed with respect to the first gate by $t_\text{e}$, during which time the state evolves freely around the Z-axis of the Bloch sphere. \textbf{b}, $P_\text{1}$ as a function of $V_\text{GL}$ and $t_\text{e}$ for states initialized near $|Y\rangle$. \textbf{c}, Line-cut of $P_\text{1}$ near $V_\text{GL}$ = -390 mV, showing $\approx$ 4.5 GHz Ramsey fringe. Red solid curve shows a fit to exponentially damped sine wave with best fit parameter $T_{2}^*$ = 1.3 ns. 
\textbf{d-e}, Two axis control of the qubit \textbf{d},  $P_\text{1}$ as a function of $\phi$ and $t_\text{e}$. \textbf{e}, Line-cut of \textbf{d} at $\phi$ = 0 (X-axis, black), 90\degree (Y-axis, red), and 180\degree (-X-axis, blue). The coherent z-axis rotation along with the rotation axis control with $\phi$  demonstrates full control of the qubit states around three orthogonal axes on the Bloch sphere. 
}
\label{fig2} 
\end{figure*}

Coherent oscillations between qubit states $\ket{0}$ and $\ket{1}$ are implemented by applying the microwave sequence shown in
Fig.~1c, using an ac excitation frequency $f_\text{ex}\approx $~4.5~GHz, which is resonant with the qubit at the sweet spot, and an ac microwave amplitude $V_\text{ac} = 70$~mV (as measured at at the arbitrary waveform generator; the amplitude at the sample is lower because of attenuation of the coaxial cables in the dilution refrigerator).
Fig.~1d shows the resulting microwave-driven Rabi oscillations in $P_{1}$, measured  as a function of the microwave burst duration $t_{\text{b}}$ and the gate voltage $V_{\text{GL}}$, which determines the base level of $\varepsilon$.
Fig.~1e  shows a line cut at $V_\text{L}= -390$~mV, corresponding to $\varepsilon \approx 0$, revealing periodic oscillations in $P_1$ at a frequency $f_\text{Rabi}\approx 1$~GHz. The dependence of $P_{1}$ on $t_\text{b}$ is well fit by an exponentially damped sine wave (Fig.~1e, red solid curve), with the best fit yielding a coherence time ${T_2}^*=1.5$~ns. 

Fig.\ 1f reports the power dependence of the qubit oscillations, showing $P_{1}$ as a function of $t_\text{b}$ and the peak-to-peak microwave amplitude $V_\text{ac}$.  The observed oscillation frequency varies linearly with $V_\text{ac}$, as expected for Rabi oscillations. Taking the filtering and attenuation in the dilution refrigerator into account, we estimate the peak-to-peak amplitude at the sample for $V_\text{ac}=120$~mV to be 1.1~mV. The maximum measured Rabi frequency of 2~GHz approaches the intrinsic frequency of the qubit (4.5 GHz) and is observed with only moderate applied power.  These oscillations correspond to $X$-rotations of the qubit on the Bloch sphere, and for the fastest rotations observed an $\text{X}_{\pi/2}$ gate requires a time of 125~ps.

We now demonstrate Z-axis rotations of the qubit through the performance of a Ramsey fringe experiment, using the microwave pulse sequence shown schematically in Fig.~2a.  We first prepare the state $\ket{Y}=\sqrt{1/2}(\ket{0}+i\ket{1})$ by initializing to $\ket{L}$, adiabatically changing the detuning to $\varepsilon$ = 0 $\mu$eV to evolve the state to $\ket{0}$, and then performing an $\text{X}_{\pi/2}$ rotation at $\varepsilon$~=~0 $\mu$eV.  Z-axis rotation results from the evolution of a relative phase between states $\ket{0}$ and $\ket{1}$, given by $\varphi={-t_\text{e}2\Delta_{1} /\hbar}$, where $t_\text{e}$ is the time spent at $\varepsilon~=~0$. The resulting state is rotated by a second $\text{X}_{\pi/2}$ microwave pulse, and the final probability $P_{1}$ again is measured by adiabatically projecting state $\ket{1}$ to $\ket{R}$ at the readout position $\varepsilon_\text{r}$. Figs.~2b and c show the resulting quantum oscillations of the qubit state around the Z-axis of the Bloch sphere and correspond to a Ramsey fringe experiment.  By fitting the oscillations to an exponentially damped sine wave (red solid curve), we extract a dephasing time $T_{2}^*= 1.3$~ns and an oscillation frequency $2\Delta/h= 4.5$~GHz, the latter of which is consistent with the spectroscopic measurements shown in Fig.~1b.

A key advantage of ac gating over dc pulsed-gating is the ability to choose the rotation axis to point in an arbitrary direction in the X-Y plane of the Bloch sphere by controlling the phase of the applied microwave burst.
On resonance in the rotating frame, the Hamilton takes the form $H = \cos(\phi)\sigma_\text{x} + \sin(\phi)\sigma_\text{y}$, where the $\sigma_i$ are Pauli matrices and $\phi$ is the relative phase of the microwave burst with respect to the first pulse incident on the qubit~\cite{Shore:1990theory}. Controlling the phase of pulses subsequent to the initial pulse, as shown in Fig.~2a, thus allows rotations of the qubit around any axis lying in the $X$-$Y$ plane of the Bloch sphere. Fig.~2d shows measurement of $P_{1}$ using such a pulse sequence as a function of both $\phi$ and $t_\text{e}$, demonstrating smooth variation in $P_{1}$ to changes in the control parameters. Fig.~2e shows line cuts of $P_{1}$ at $\phi$=0, 90$\degree$, and 180$\degree$ (corresponding to the second pulse inducing a $\pi/2$ rotation around the $X$, $Y$, and $-X$ axes, respectively); microwave phase control clearly enables control of the phase of the resulting Ramsey fringes.  Taken together, the data summarized in Figs.~1 and 2 demonstrate control of the qubit over the entire Bloch sphere, including both $Z$-axis rotations and rotations about any arbitrary axis lying in the $X$-$Y$ plane.

\begin{figure}
\includegraphics[width=0.47\textwidth]{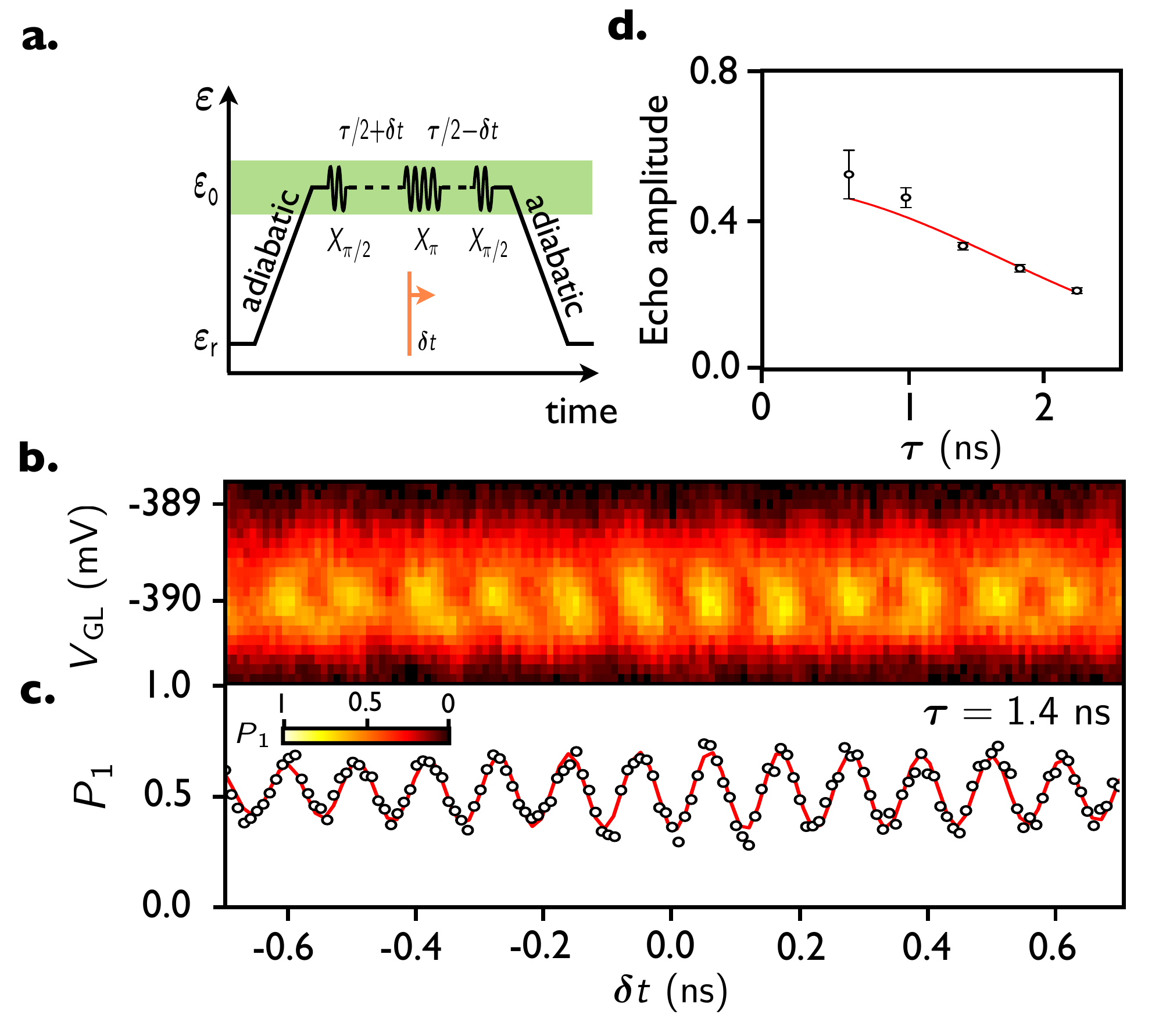}
\caption{
\textbf{Hahn echo measurement.} \textbf{a}, Schematic pulse sequence for the measurement of Hahn echo that corrects for noise that is static on the time scale of the pulse sequence~\cite{Koppens:2008p236802, Vanderypen:2005p1037, Dial:2013p146804}.
\textbf{b-c}, Typical echo measurement with fixed total evolution time
$\tau = 1.4$~ ns. \textbf{b}, $P_{1}$ as a function of $V_{GL}$ and delay time $\delta t$ of the X$_{\pi}$ pulse with decoupling $X_\pi$ gate. The effects of static inhomogeneities are minimized at $\delta t=0$, and oscillations of $P_1$ as function of $\delta t$ at twice the Ramsey frequency decay with $\delta t$ at the inhomogeneous decay rate $1/T_2^{*}$. The magnitude of the signal at $\delta t=0$ as the wait time $\tau$ is varied decays at the
homogeneous decay rate $1/T_2$.
\textbf{c}, Line cut of $P_{1}$ near  $V_{\text{GL}}$ = -390\ mV showing oscillatory signal at twice the Ramsey frequency, $\approx$ 9 GHz. Solid red curve is a fit to gaussian envelope with fixed $T_{2}^*$ = 1.3 ns, consistent with the Ramsey fringe measurement. 
\textbf{d}, Echo amplitude as a function of $\tau$. The solid red curve is a gaussian fit with $T_{2,\text{echo}}$ = 2.1 ns. Applying the Hahn echo sequence increases the dephasing time, indicating that a significant component of the dephasing arises from low-frequency noise processes.
}
\label{fig3} 
\end{figure}

We characterized decoherence times by implementing a Hahn echo~\cite{Koppens:2008p236802, Vanderypen:2005p1037, Dial:2013p146804} of the ac-gated charge qubit by applying the pulse sequence shown in Fig.\ \ref{fig3}a. Inserting an $X_{\pi}$ pulse between  state initialization and measurement, which is performed with $X_{\pi/2}$ gates, corrects for noise that is static on the time scale of the pulse sequence. In Fig.\ \ref{fig3}b and c, while keeping the total free evolution time $\tau$ fixed, we sweep the position of the decoupling $X_{\pi}$ pulse to reveal an echo envelope~\cite{Dial:2013p146804, Shi:2013p075416}. 
The maximum amplitude of the observed envelope reveals the extent to which the state has dephased during the free evolution time $\tau$, characterized by $T_2$, whereas the amplitude decays as a function of $\delta t$ with inhomogeneous decay time $T_2^{*}$. The oscillations of $P_{1}$ in Fig.\ \ref{fig3}b and c  are observed as a function of $\delta t$ at twice the Ramsey frequency ($2f_\mathrm{Ramsey}\approx 9$~GHz) and are well-fit by a Gaussian decay (red solid curve). Fig.\ \ref{fig3}d shows the echo amplitude decay as a function of $\tau$, where for each $\tau$ the echo amplitude is determined by fitting the echo envelop to Gaussian decay similar to Fig.\ \ref{fig3}c. By fitting the echo amplitude decays to a Gaussian, we obtain the dephasing time $T_\text{2}\approx 2.2$~ns.

\begin{figure*}
\includegraphics[width=0.95\textwidth]{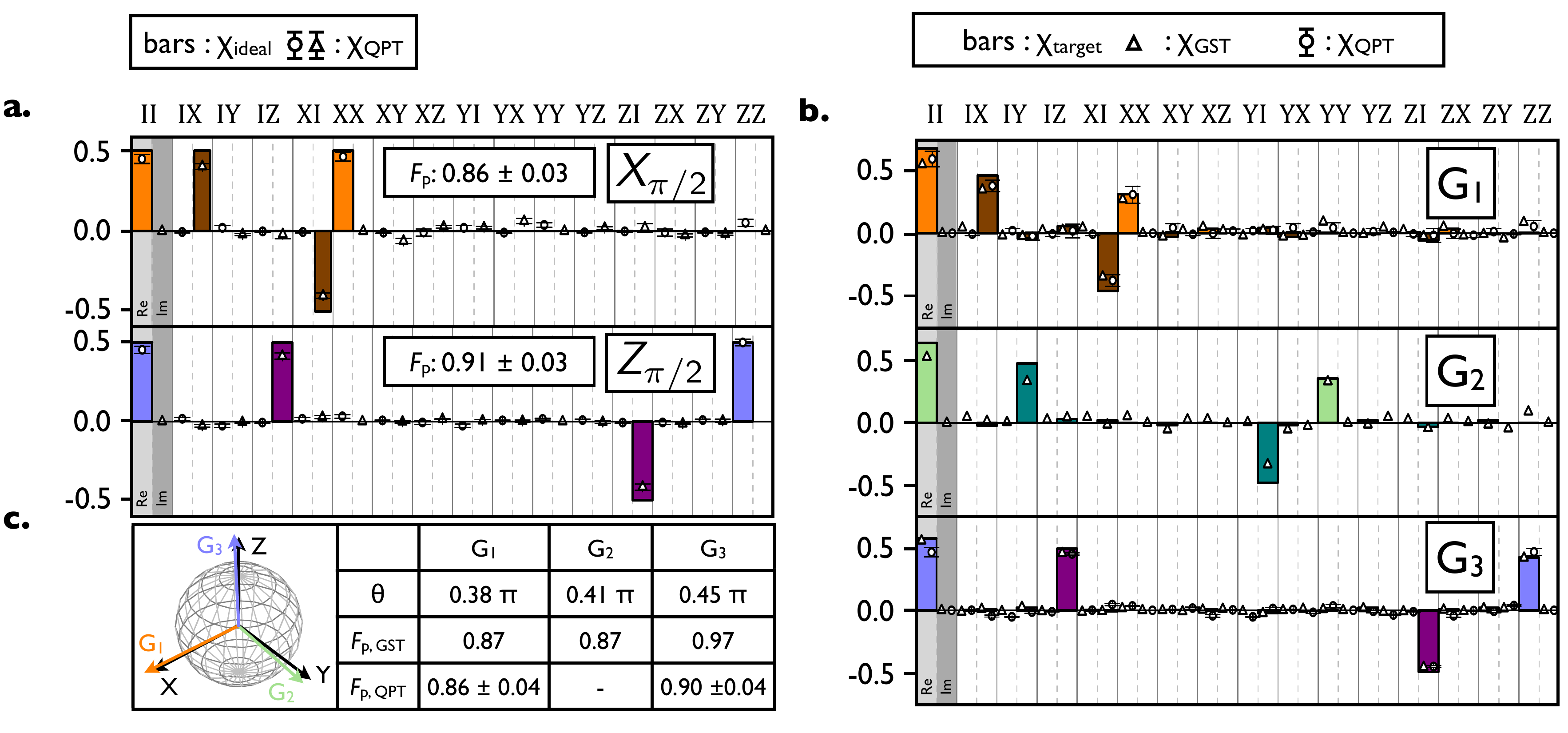}
\caption{
\textbf{Quantum process tomography and gate set tomography of the ac-gated charge qubit. a}, 
Real and imaginary parts of the elements of the process matrix $\chi$~\cite{Nielsen:2000} in the Pauli basis $\{$\text{I}$, X, Y, Z\}$ for $X_{\pi/2}$ and $Z_{\pi/2}$ processes : ideal ``targets" (solid bars), and standard QPT estimates (open circles).
\textbf{b},  $\chi$ for the uncalibrated operations $G_{1}$, $G_{2}$, and $G_{3}$, obtained by gate set tomography~\cite{BlumeKohout:2013p1310.4492} (GST, triangles) and standard quantum process tomography~\cite{Nielsen:2000,Chow:2009p090502, Kim:2014nature, Chow:2009p090502} (QPT, open circles), compared to target gates $T_{1}$, $T_{2}$, and $T_{3}$ (solid bars). Since these gates are not precalibrated, the target gates are defined to be the unitary processes closest to the GST estimates of $G_{1}$, $G_{2}$, and $G_{3}$ in Frobenius norm.  GST self-consistently determines the state preparation, gate operations, and measurement processes~\cite{BlumeKohout:2013p1310.4492}.
\textbf{c}, Rotation axes on the Bloch sphere, rotation angle $\theta$, process fidelities obtained by GST ($F_{\text{p,GST}}$), and QPT ($F_{\text{p,QPT}}$)  for three processes $G_{1}$, $G_{2}$, and $G_{3}$. Here, the rotation axis and angle correspond to the closest unitary operations to the GST estimate ($T_{1}$, $T_{2}$, and $T_{3})$; 
the process fidelities are also taken between the  estimates and these target processes. The error on $F_{\text{p,QPT}}$ was estimated by repeating QPT using 10 distinct sets of input and output states; standard deviations are reported. GST and QPT yield consistent results, with process fidelities  $\geq86\%$ for all gates.}
\label{fig4} 
\end{figure*}

The ultimate test of experimental qubit control is the demonstration of repeatable quantum logic gates. Although $\pi/2$ rotations that generate the Clifford group are commonly demonstrated, ac control allows direct implementation of any unitary.  We therefore demonstrated (and validated with quantum tomography) two distinct gatesets:  (1) high-fidelity approximations to $\{X_{\pi/2},Z_{\pi/2}\}$; and (2) a set of three arbitrarily chosen near-unitary operations $\{G_1,G_2,G_3\}$.  We used standard quantum process tomography (QPT)~\cite{Nielsen:2000,Chow:2009p090502} to characterize the first gateset.  Fig.~\ref{fig4}(a) shows the resulting process matrices ($\chi$) expressed in the Pauli basis: solid bars represent the ideal ``target" quantum processes, while open circles show the results of QPT.  The process fidelities \cite{MagesanPRA11} between the QPT estimates and the targets are $F=86\%$ and $F=91\%$ for the $X_{\text{$\pi/2$}}$ and $Z_{\text{$\pi/2$}}$ operations, respectively.
QPT analysis is not entirely reliable, because QPT relies on prior knowledge of input states and final measurements that are implemented using the same logic gates that we seek to characterize (see Supplementary Information S3).  Therefore, we also applied a technique called \emph{gate set tomography} (GST)~\cite{BlumeKohout:2013p1310.4492} that avoids these assumptions.  GST characterizes logic gates and state preparation/measurement simultaneously and self-consistently, by representing all of them as unknown process matrices.  Because this frees us from any obligation to use carefully calibrated operations, we applied GST to a set of three repeatable but uncalibrated logic gates that we denote $\{G_1,G_2,G_3\}$.

Data for GST are obtained from many repetitions of several specific experiments, each described by a specific sequence of operations: (i) initialize the qubit in state $\rho$; (ii) perform a sequence of $L\in[0\ldots32]$ operations chosen from $\{G_1,G_2,G_3\}$; (ii) perform measurement $\mathcal{M}$.  Statistical analysis (a variant of maximum likelihood estimation) is used to find the estimates $\{\hat{G}_1,\hat{G}_2,\hat{G}_3,\rho,\mathcal{M}\}$ that are most consistent with the 
measurements
(see the Supplementary Information S3 for more details).  Since we did not set out to implement any particular rotations, we compute \emph{ex post facto} the closest unitary rotations to these estimates, and define these as the ``target" gates.  The results are shown in Figs.\ref{fig4}b-c.  Fig.\ref{fig4}b shows the elements of the process ($\chi$) matrix for the GST estimates (triangles) and the closest-unitary ``targets" (solid bars).  Fig.\ref{fig4}c portrays those same unitaries as rotations of the Bloch sphere.

We also performed QPT on the $G_1$ and $G_3$ gates to confirm and validate the GST results.  QPT depends critically on known gates (used to prepare known input states and perform known measurements), so we used the GST closest-unitary approximations for $G_1$ and $G_3$ to model the preparation of input states and final measurements for QPT ($G_2$ was neither necessary nor used).  The results, shown as open circles in in Fig.\ref{fig4}b, are consistent with the GST results. To distill a single figure of merit, we computed the process fidelity $F$ between our estimates (both GST and QPT) and the closest unitary rotation.  Its interpretation is slightly different in this context; since the ``targets" were computed \emph{ex post facto}, $1-F$ quantifies the amount of \emph{incoherent} error in our gate implementation (whereas with a pre-existing target, it also quantifies coherent under/over-rotation errors).  These process fidelities, shown for both GST and QPT in Fig.\ref{fig4}c, are consistent both with each other and the process fidelities calculated in Fig.\ref{fig4}a.

Coherent microwave ac-gating of a semiconductor quantum dot charge qubit offers fast ( $>$GHz) manipulation rates for all elementary rotation axes. Moreover, because all gates can be performed at the sweet spot where the decoherence time is in the order of nanoseconds instead of $\sim$ 100~ps~\cite{Dovzhenko:2011p161802,Shi:2013p075416}, rotations around three orthogonal axes of the Bloch sphere with process fidelities higher than 86\% are achieved. This improvement is analogous to the early development in the superconducting charge qubits ~\cite{Nakamura:1999p786, Vion:2002p886}, where operating at sweet spot (quantronium) with resonant microwaves ~\cite{Vion:2002p886} demonstrated the first high quality universal single qubit gate operations after initial demonstration of charge qubit manipulation with non-adiabatic pulse techniques ~\cite{Nakamura:1999p786}. Applying Hahn-echo decoupling sequence provides modest improvement in coherence time ($T_\text{2,echo} \approx$ 2.1  ns compared to $T_{2}$* $\approx$ 1.3 ns), indicating that understanding high frequency charge noise as well as charge relaxation at the sweet spot will be important for further development. The quantum dot charge qubit is highly tunable using gate voltages, and we expect that investigating coherence and process fidelity as a function of tunnel coupling strength between the dots will provide an effective route to improve its performance~\cite{Petersson:2010p246804}. 

\vspace{.2in}
\noindent \textbf{Methods}

\emph{Measurement:} 
The experiments are performed on a double quantum dot fabricated in a Si/SiGe heterostructure~\cite{Shi:2013p075416,Simmons:2011p156804} at base temperature (electron temperature $\simeq 140$~mK~\cite{Simmons:2011p156804}) in a dilution refrigerator. The valence electron occupation of the double dot is $(2,1)$ or $(1,2)$, as confirmed by magnetospectroscopy measurements~\cite{Simmons:2011p156804}. All manipulation sequences including microwave bursts are generated by a Tektronix 70002A arbitrary waveform generator and are added to the dot-defining dc voltage through a bias tee (Picosecond Pulselabs~5546-107) before being  applied to gate GR. Similarly to our previous study\ \cite{Kim:2014nature}, the conductance change through the quantum point contact (QPC) with and without the applied microwave burst, and measured with a lock-in amplifier (EG\&G model~7265), is used to determine the average charge occupation and is converted to the reported probabilities. Charge relaxation during the measurement phase is taken into account using the measured charge relaxation time $T_1\simeq 23.5$~ns at the read-out detuning of $\varepsilon_\text{R}\simeq -160$~$\mu$eV \cite{Kim:2014nature}. In the Supplementary Information S1, we present the details of measurement technique and the probability normalization.  

\emph{Gate set tomography (GST)}:
GST is performed in two stages, as described in \cite{BlumeKohout:2013p1310.4492}. 
First, data obtained from short gate sequences are analyzed using linear inversion, to obtain a rough estimate.  Next, data from long sequences are incorporated using maximum-likelihood parameter estimation, to refine the preliminary estimate.  The required gate sequences are defined using a set of \emph{fiducial sequences} $\mathcal F=\{F_i\}$ that fix a consistent (but \emph{a priori} unknown) reference frame.  For this experiment, we chose $\mathcal F = \{\emptyset,G_1,G_2,G_1^3 \}$, where $\emptyset$ is the null operation (do nothing for no time).  In terms of these fiducials, linear inversion GST demands data from all sequences of the form $F_iG_kF_j$ where $i,j=1\ldots4$ and $k=0\ldots 3$ and $G_0\equiv\emptyset$ is the null operation. Supplementary Information S3 provides more details about the GST implemented here.

\vspace{.2in}

\emph{Acknowledgements}
This work was supported in part by ARO (W911NF-12-0607), NSF(DMR-1206915), NSF(PHY-1104660), and by the Laboratory Directed Research and Development program at Sandia National Laboratories.  Sandia National Laboratories is a multi-program laboratory managed and operated by Sandia Corporation, a wholly owned subsidiary of Lockheed Martin Corporation, for the U.S. Department of Energy's National Nuclear Security Administration under contract DE-AC04-94AL85000.  Development and maintenance of the growth facilities used for fabricating samples is supported by DOE (DE-FG02-03ER46028). This research utilized NSF-supported shared facilities at the University of Wisconsin-Madison.

\emph{Author Contributions}
DK performed electrical measurements, state and process tomography, and analyzed the data with MAE, MF and SNC. DRW developed hardware and software for the measurements. CBS fabricated the quantum dot device. JKG, RB-K, and EN performed gate-set tomography. DES and MGL prepared the Si/SiGe heterostructure. All authors contributed to the preparation of the manuscript. 

\emph{Additional Information}
Supplementary information accompanies this paper. Correspondence and requests for materials should be addressed to Mark A. Eriksson (maeriksson\emph{@}wisc.edu).


\renewcommand{\theequation}{S\arabic{equation}}
\setcounter{equation}{0}
\renewcommand{\thefigure}{S\arabic{figure}}
\setcounter{figure}{0}
\renewcommand{\thesection}{S\arabic{section}}
\setcounter{section}{0}

\section*{Supplementary Information}

\section{Details of time-averaged measurement and probability normalization}
\label{sup:measurement}
\begin{figure}[b]
\includegraphics[width=0.47\textwidth]{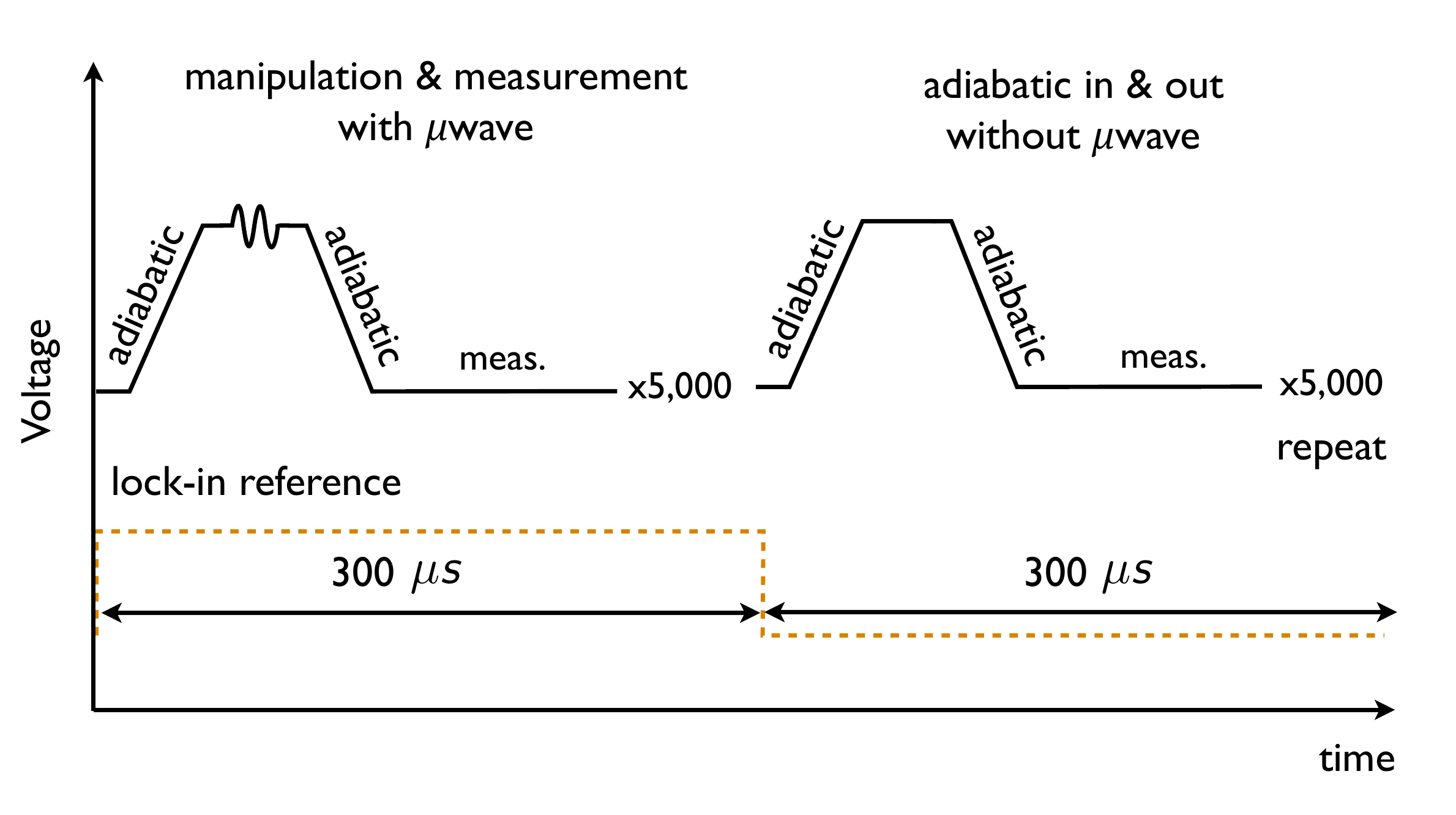}
\caption{
Pulse sequence used for lock-in measurement of the qubit state. The orange dashed line shows the corresponding lock-in reference signal, which also serves as the probability calibration pulse.  }
\label{fig:S1} 
\end{figure}

For the measurement of changes in the probabilities of charge occupation resulting from fast microwave bursts, we use the general approach described in~\cite{Kim:2014nature}, where we measure the difference between the QPC conductance with and without the manipulation pulse train. Fig.~\ref{fig:S1} shows similar scheme adopted in this work. We alternate an appropriate number of manipulation and measurement sequences (measurement time $\sim$60~ns) with microwave bursts with adiabatic in and out sequences without microwave to form a low frequency signal with frequency on the order of $\sim$1~kHz. The adiabatic ramp sequence without microwave manipulation does not induce state population change, but greatly reduces background signal due to capacitive crosstalk. The manipulation sequence including the microwave bursts is generated using Tektronix AWG70002A arbitrary waveform generator with maximum sample rate of 25 Gs/s and analog bandwidth of about 13 GHz. The data are acquired using a lock-in amplifier with a reference signal corresponding to the presence and absence of the pulses, as shown schematically by the orange dashed line in Fig.~\ref{fig:S1}. We compare the measured signal level with the corresponding $(2,1)$-$(1,2)$ charge transition signal level, calibrated by sweeping gate GL and applying the orange square pulse shown in Fig.~\ref{fig:S1} to gate GL.  Similarly to  previous work~\cite{Kim:2014nature},  charge relaxation during the measurement phase is taken into account using the measured charge relaxation time $T_1\approx 23.5$~ns at the read-out detuning of $\varepsilon_{r}\approx -160$~$\mu$eV. 

\section{Model Hamiltonian corresponding to Fig. 1c}
The energy level diagram is calculated from the Hamiltonian written in the basis of the ground and first excited states of the electron charge states $(2,1)~=~\ket{L}$ and $(1,2)~=~\ket{R}$: $\{ \ket{L}_\text{g} , \ket{L}_\text{e}, \ket{R}_\text{g}, \ket{R}_\text{e}\}$.  In this basis, $H$ is given by 
\begin{equation}
\label{eq:S3}
H=
\begin{pmatrix}
\varepsilon/2 & 0 & \Delta_1 & -\Delta_2 \\
0 & \varepsilon/2+\delta E_\text{L} & -\Delta_3 & \Delta_4 \\
\Delta_1 & -\Delta_3 & -\varepsilon/2 & 0 \\
-\Delta_2 & \Delta_4 & 0 & -\varepsilon/2+\delta E
\end{pmatrix} .
\end{equation}
Here, $\Delta_{1-4}$ are tunnel coupling matrix elements, $\varepsilon$ is the detuning, and $\delta E_\text{L}$ and $\delta E$ are the energy separation of the ground and excited $\ket{L}$ and $\ket{R}$ states, respectively. The parameters $\Delta_3$, $\Delta_4$, and $\delta E_\text{L}$ are relevant to high energy states which we do not experimentally access in this work, and we use the values determined from our previous study~\cite{Shi:2014p3020}. We used parameters of tunnel coupling between $|L\rangle$ and $|R\rangle$ $2\Delta_{1}/h$ = 5.2 GHz, tunnel coupling between $|L\rangle$ and low lying excited state $\ket{R}_\text{e}$ on the right dot $2\Delta_{2}/h$ = 14.5 GHz, and singlet-triplet energy splitting on the right dot $\delta R$ = 12.1 GHz, extracted from measurements as in Refs.~\cite{Shi:2014p3020,Kim:2014nature}.

\section{State and process tomography}
\label{sup:tomography}

\subsection{Details of the linear-inversion gate set tomography experiments}
Gate set tomography (GST) is a method for self-consistently characterizing state preparation, measurement, and quantum operations.  Our method is described in detail in Ref.~\cite{BlumeKohout:2013p1310.4492} (Ref.~\cite{MerkelPRA13} demonstrated a similar protocol); this section discusses some specific issues encountered in this experiment.

Like standard QPT, GST requires an \emph{informationally complete} set of measurements -- i.e., at least one linearly independent probability per gateset parameter.  But it eliminates standard QPT's need for known initialization and measurement processes by doing initialization and measurement with unknown operations selected from the \emph{gateset} to be characterized.  Informational completeness is achieved by using fiducial sequences ($\mathcal{F}=\{F_i\}$) built from the gates themselves to generate at least $d^2=4$ distinct input states, and to rotate the final state into at least $d^2=4$ distinct measurement bases.  However, whereas in standard QPT informational completeness can be ensured in advance, in GST we do not know in advance what operations the $F_i$ will perform.  Instead, we test the \emph{data} for the signature of informational completeness, and if necessary adjust the fiducial sequences.  This signature is the spectrum of the Gram matrix $G$ defined by $G_{i,j} = P_{1}(F_{i}F_{j})$.  $G$ should have rank $d^2$ if the fiducials $\mathcal{F}$ are informationally complete for a $d$-dimensional system.  In the experiments reporte here, our initial choice of $\mathcal{F}$ accidentally produced Gram matrix with only three significant eigenvalues; we adjusted the fiducial sequences to ensure that all four eigenvalues of $G$ were large ($\geq$ 0.19).
With our initial microwave-pulse gates, chosen to roughly approximate $\pi/2$ rotations, the fiducial set $\mathcal{F} = \{\emptyset, G_{1}, G_{2}, {G_{1}}^2\}$ yielded a Gram matrix with smallest eigenvalue of 0.05, which caused the linear inversion GST to be ill-conditioned. Changing the fourth fiducial to $F_4 = G_1^3$ as well as reducing the rotation angle by changing the amplitude of the ac driving yielded a Gram matrix with smallest eigenvalue 0.19, which ensures reliable linear inversion for the first step of GST. As is shown in Fig.~4c of the main text, GST estimates the rotation angles of $G_{1}$, $G_{2}$, and $G_{3}$ to be $0.38\pi$, $0.41\pi$, and $0.45\pi$, respectively.

\subsection{Details of the non-linear gate set tomography refinement}
GST proceeds in two steps: a linear inversion step, where we obtain coarse estimates of the gates, followed by a refinement step.
Higher accuracy is attained by performing longer sequences of gates.  In this experiment, we performed (and estimated the measurement probability associated with) sequences of the form $F_i G_k^n F_j$, for $i,j=1\ldots4$, $k=1\ldots 3$, and $n=2\ldots32$.  Many repetitions amplify over/under-rotation error (e.g. by an angle $\theta$) in each gate $G_k$, so that by measuring $n\theta$ to reasonable accuracy we achieve a very accurate estimate of $\theta$.  However, because the associated measurement probabilities are highly nonlinear in $G_k$, we cannot use linear inversion.
Instead, we estimated the gateset $\mathcal{G} = \{\hat{G}_k,\hat{\rho},\hat{\mathcal{M}}\}$ by maximizing the likelihood $\mathcal{L}(\mathcal{G}) = \mathrm{Pr}(\mathrm{data}|\mathcal{G})$.  Since the likelihood function $\mathcal{L}(\mathcal{G})$ is in general nonconvex, we used the linear inversion estimate as a starting point for local optimization.  We found that in this case the best local optimization procedure (resulting in the highest maximum likelihood) was to repeatedly apply the Nelder-Mead downhill simplex method until it converged.  Since the resulting estimate is not necessarily a physically valid gateset (those for which $\rho$ and $\mathcal{M}$ are valid quantum states/measurements, and each $G_k$ is a completely positive, trace-preserving map), it is projected onto the space of valid gatesets to produce a final estimate.  This final maximum-likelihood estimate predicts count statistics that fit the observed data well, but the variability of results with different local optimization procedures indicates that the likelihood contains a number of local minima and suggests that even in our best case the likelihood may not have been strictly maximized.
However, this effect appears to be less significant than non-Markovian noise effects, which are not accounted for in GST (unlike QPT, GST can incorporate data from experimental sequences in which a single gate $G_k$ appears more than once -- but it does so by explicitly assuming that $G_k$ applies the same quantum process each time it is applied, with no systematic variation with respect to time or context).

\subsection{Standard quantum process tomography (QPT)}
We perform standard process tomography on our qubit in the usual way, by preparing precalibrated, informationally complete states, applying the unknown operation to be characterized, and measuring in a precalibrated, informationally complete basis.
In our experiment, the state preparation and measurement processes are precalibrated using Rabi oscillation and Ramsey fringe experiments, as detailed in the main text.
We perform QPT to characterize a total of four processes: $X_{\text{$\pi/2$}}$, $Z_{\text{$\pi/2$}}$ (shown in Fig.~4a in the main text), and $G_1$ and $G_3$ (shown in Fig.~4b in the main text).
In all cases, first we initialize to the $\left| L \right>$ state, then adiabatically transition to the $\left| 0 \right>$ state, as detailed in the main text.
Next, we prepare an informationally complete set of inputs in two separate ways. First, for the $X_{\text{$\pi/2$}}$ and $G_1$ (which is $X$-like) operations, we apply the operations $X_{\pi/2}Z_{\pi/2}$ and   $X_{\Phi}$ for three separate angles $\Phi$. 
This gives the $\ket{-X}$ state, in addition to three states that lie on the $y-z$ plane, forming an informationally complete set.
Second, for the $Z_{\text{$\pi/2$}}$ and $G_3$ (which is $Z$-like) operations, we either do nothing or apply the operations $Z_{\Phi}X_{\pi/2}$ for three separate angles $\Phi$. This gives for our inputs the states $\ket{0}$ and three states in the $x-y$ plane, which are again informationally complete. The reference time which sets $\Phi$ is varied in the experiment and the results with 10 different $\Phi$ were used to estimate statistical average and standard deviation of the process matrices and fideltities. We choose these two different input sets for X or Z-gate process tomography for data post-processing convenience. After preparing the informationally complete input set, we then implement the processes to be characterized. Finally, we perform $-Y_{\pi/2}$, $X_{\pi/2}$, and identity operations, followed by a measurement in the $Z$-basis. Note that in the state tomography of X-gate we implement tomography in the rotating frame by adjusting phase of the measurement $\pi/2$ pulses taking phase accumulation during manipulation pulse into account, whereas Z-gate tomography is performed in the lab frame where the phase of the measurement pulses are fixed.
Once these measurements are complete, we use maximum-likelihood estimation~\cite{NielsenBook,Chow:2009p090502, Kim:2014nature} to reconstruct the  $\chi$ matrix representation of the unknown process, given by~\cite{NielsenBook,Chow:2009p090502}
\begin{equation} 
{\cal E}(\rho)=\sum_{m,n=1}^{4}\tilde{E}_m\rho\tilde{E}^\dagger_n\chi_{mn} ,
\end{equation}
where ${\cal E}(\rho)$ is the density matrix specifying the output for a given input density matrix $\rho$, $\tilde{E}_m$ are the basis operators in the space of $2\times 2$ matrices, and $\chi$ is the process matrix.
The entries of this matrix are plotted in Fig.4a of the main text.

\bibliography{siliconqcsnc}

\end{document}